\documentclass[aps,prl,twocolumn,amsmath,amssymb,nofootinbib,superscriptaddress]{revtex4}

\newcommand{\bra}[1]{\langle#1|}
\newcommand{\ket}[1]{|#1\rangle}

\usepackage[pdftex]{graphicx}
\usepackage{mathrsfs}
\usepackage[colorlinks]{hyperref}

\begin{document}

\bibliographystyle{apsrev}

\title{Will boson-sampling ever disprove the Extended Church-Turing thesis?}

\author{Peter P. Rohde}
\affiliation{Centre for Engineered Quantum Systems, Department of Physics \& Astronomy, Macquarie University, Sydney NSW 2113, Australia}
\email[]{dr.rohde@gmail.com}
\homepage{http://www.peterrohde.org}

\author{Keith R. Motes}
\affiliation{Centre for Engineered Quantum Systems, Department of Physics \& Astronomy, Macquarie University, Sydney NSW 2113, Australia}

\author{Paul A. Knott}
\affiliation{NTT Basic Research Laboratories, NTT Corporation, Atsugi, Kanagawa 243-0198, Japan}

\author{William J. Munro}
\affiliation{NTT Basic Research Laboratories, NTT Corporation, Atsugi, Kanagawa 243-0198, Japan}

\date{\today}

\frenchspacing

\begin{abstract}
Boson-sampling is a highly simplified, but non-universal, approach to implementing optical quantum computation. It was shown by Aaronson \& Arkhipov that this protocol cannot be efficiently classically simulated unless the polynomial hierarchy collapses, which would be a shocking result in computational complexity theory. Based on this, numerous authors have made the claim that experimental boson-sampling would provide evidence against, or disprove, the Extended Church-Turing thesis -- that any physically realisable system can be efficiently simulated on a Turing machine. We argue against this claim on the basis that, under a general, physically realistic independent error model, boson-sampling does not implement a provably hard computational problem in the asymptotic limit of large systems.
\end{abstract}

\maketitle

Boson-sampling \cite{bib:AaronsonArkhipov10} has been presented as a new, highly simplified, yet limited form of linear optics quantum computation \cite{bib:KLM01}. It has attracted interest because, despite not being universal for quantum computation, it implements a classically hard algorithm using far fewer physical resources than conventional approaches.

In the boson-sampling model we begin by preparing a multi-mode Fock state, comprising $n$ single photons in $m$ modes,
\begin{equation}
\ket{\psi_\mathrm{in}} = \ket{1_1,\dots,1_n,0_{n+1},\dots,0_m}.
\end{equation}
This input state is passed through a linear optics network, comprising only beamsplitters and phase-shifters, implementing a unitary transformation $\hat{U}$. Importantly, there is no active feedforward, measurement or quantum memory within the circuit, representing a significant simplification compared to universal linear optics quantum computing schemes \cite{bib:KLM01}. The output state to the interferometer is a superposition of all possible photon-number configurations, subject to the constraint that total photon number is preserved,
\begin{equation}
\ket{\psi_\mathrm{out}} = \hat{U} \ket{\psi_\mathrm{in}} = \sum_S \gamma_S \ket{n_1^{(S)},\dots,n_m^{(S)}},
\end{equation}
where $S$, of which there are an exponential number, are the different photon number configurations and $n_i^{(S)}$ is the number of photons in the $i$th mode associated with configuration $S$. Following the interferometer we perform an $m$-fold number-resolved coincidence measurement of the output modes. We repeat the experiment many times, each time sampling from the probability distribution \mbox{$P(S)=|\gamma_S|^2$}. It was shown by Aaronson \& Arkhipov \cite{bib:AaronsonArkhipov10} that this sampling problem is likely classically hard to simulate, offering an exponential quantum speed-up compared to the best known classical algorithm. The presumed classical hardness of this problem relates to (1) there are an exponential number of terms in the output superposition, and (2) each of the amplitudes $\gamma_S$ is proportional to a (different) matrix permanent, which resides in the complexity class \#\textbf{P}-complete, a class higher in the complexity hierarchy than \textbf{NP}-complete, which is presumed to be classically hard to calculate. Note that boson-sampling does not actually let us \emph{calculate} matrix permanents as this would require an exponential number of measurements.

The presumed classical hardness of boson-sampling, combined with its relative simplicity, has attracted much interest from experimentalists, who wish to demonstrate architectures with quantum speed-up using the fewest possible physical resources. Recently, there have been four demonstrations of boson-sampling using three photons \cite{bib:Broome20122012, bib:Crespi3, bib:Tillmann4, bib:Spring2}.

Any physical system will exhibit imperfections, however small. Let us assume that each of the desired input single photon states is actually the desired single photon state with probability $p$, otherwise some residual state comprising erroneous terms, which we will label $\hat\rho_\mathrm{error}$. Then our input state is of the form,
\begin{equation}
\hat\rho_\mathrm{in} =\left(\bigotimes_{i=1}^n[p\ket{1}\bra{1} + (1-p)\hat\rho_\mathrm{error}^{(i)}]\right) \otimes [\ket{0}\bra{0}]^{\otimes^{m-n}},
\end{equation}
where, for generality, $\hat\rho_\mathrm{error}^{(i)}$  may be distinct for each input mode $i$. This independent error model is very general and applies to a variety of physically realistic errors. In the case of photon-number errors, $\hat\rho_\mathrm{error}$ collects all non-single-photon terms. For example, if the photon source has a probability of loss and a probability of second order excitation, then \mbox{$\hat\rho_\mathrm{error} = p_0 \ket{0}\bra{0} + p_2\ket{2}\bra{2}$}. Similarly, spectral impurity and mode-mismatch are common errors in optical quantum computing. In this instance we can let $\ket{1}$ correspond to a desired spectral mode, which overlaps with all other photons, whilst $\hat\rho_\mathrm{error}$ collects spectral components orthogonal to the desired mode. In either case, $p$ quantifies how close our single photons are to the desired state. We desire that \mbox{$p=1$}. But in any physically realistic system \mbox{$p<1$}.

After passing this state through the interferometer and sampling the output distribution, sometimes we will have sampled from a distribution whereby the input state was the desired \mbox{$\ket{1}^{\otimes n}\otimes \ket{0}^{\otimes m-n}$}, otherwise we have sampled from an erroneous distribution. The probability that we have sampled from the desired distribution is \mbox{$P=p^n$}. That is, the probability of sampling from the desired distribution scales inverse exponentially with the size of the system. This scaling characteristic was noted by Aaronson \& Arkhipov \cite{bib:AaronsonArkhipov10, bib:AA13response} and represents one of the major challenges facing large-scale demonstrations.

The original proof by Aaronson \& Arkhipov only considered the regime where \mbox{$P>1/\mathrm{poly}(n)$}. This bound has to-date not been loosened. One cannot rule out that loosening of this bound will be achieved in the future. However, based on present understanding, boson-sampling is only known to be hard in this regime. Thus, based on the presently best known bound, an independent error model takes us outside the regime whereby boson-sampling is known to be classically hard. Future developments in our understanding of this bound may mitigate the concerns raised here.

%It was shown by Aaronson \& Arkhipov that boson-sampling is only provably hard when \mbox{$P>1/\mathrm{poly}(n)$}. If this inequality is not satisfied, boson-sampling is not implementing a provably hard algorithm. However, \mbox{$P=p^n$}, where \mbox{$p<1$}. Thus we require \mbox{$p^n>1/\mathrm{poly}(n)$}, which cannot be satisfied in the limit of large $n$ for any non-unit value of $p$. Therefore, in the asymptotic limit of large-scale devices, any physically realistic error model will undermine the computational complexity of boson-sampling. 

It has been widely claimed that experimental demonstration of boson-sampling would provide evidence against \cite{bib:AA13response, bib:Shchesnovich13}, strongly contradict \cite{bib:Broome20122012}, or disprove \cite{bib:ShenDuan13, bib:Molmer13, bib:shchesnovich2014conditions} the Extended Church-Turing thesis (ECT) -- that any physical system can be \emph{efficiently} simulated by a Turing machine. However, the validity of the ECT thesis is an asymptotic statement -- we cannot make statements about computational complexity other than in the asymptotic limit. Since boson-sampling fails in the asymptotic limit under the physically realistic error model we presented, it is questionable to make the claim that boson-sampling provides evidence against the ECT thesis.

This argument is reminiscent of the \textbf{P} vs. \textbf{NP} debate. It has long been known that \emph{perfect} analog computers can solve \textbf{NP}-complete problems in polynomial time. Thus, one might expect that demonstration of an analog computer might provide evidence that \mbox{$\mathbf{P}=\mathbf{NP}$}. However, upon closer examination, we find that if physically realistic error models are incorporated into analog computation, their computational complexity collapses and they are only able to solve problems in \textbf{P}. Thus, for \emph{physical} rather than \emph{mathematical} reasons, analog computation will never provide evidence in the \textbf{P} vs. \textbf{NP} debate. Similarly, whilst it is mathematically rigorous that ideal boson-sampling implements a classically hard problem, once \emph{physical} effects are taken into consideration, we find that large-scale boson-sampling fails, providing no elucidation on the ECT debate.

To overcome the discussed problems we need some mechanism to accommodate for errors. In conventional quantum computing, quantum error correction and fault-tolerant codes are employed to allow for correct operation of quantum circuits even in the asymptotic limit. These techniques require intermediate measurement and feed-forward to progressively keep errors in check. In the boson-sampling model, however, we are explicitly forbidden from doing this -- we are only allowed passive operations prior to photo-detection. Instead, experimentalists employ another technique -- post-selection. Here, we simply run the device many times and throw away measurement results that are incompatible with the hypothesis. For example, to overcome spectral impurity, we perform narrowband filtering prior to photo-detection. This projects the photons' wavepackets onto frequency eigenstates, which are effectively indistinguishable, thereby keeping only the component of the output state for which the photons are indistinguishable. However, narrowband filtering in effect discards most of the photons' wavepackets, significantly reducing the effective detection efficiency. If this effective efficiency is $\eta$ then the total success probability of the device is $\eta^n$, which is exponentially small. Thus, filtering trades one exponential dependence for another, resulting in a device which is still exponentially susceptible to errors. In the case of photon-number errors, post-selection doesn't help at all. If \mbox{$\hat\rho_\mathrm{error} = p_0 \ket{0}\bra{0} + p_2\ket{2}\bra{2}$}, then post-selecting on detecting exactly $n$ photons at the output does not project us uniquely onto the $\ket{1,1,1,1,\dots}$ input state, but could have equally well projected us onto the \mbox{$\ket{2,0,2,0,\dots}$} or \mbox{$\ket{0,0,2,2,\dots}$} input states and so on, and we have ambiguity as to which distribution we were sampling from, yielding the same exponential error dependence as before.

There are two clear ways out of this conundrum. First, if it were shown that the requirement for sampling from the correct distribution scaled as \mbox{$P>1/\mathrm{exp}(n)$} rather than as \mbox{$P>1/\mathrm{poly}(n)$} this would overcome this obstacle. However, this would be an unexpected and shocking result, as it would imply that there exists a subset of quantum computation, which implements a classically hard problem, in the absence of \emph{any} kind of fault-tolerance. Were this shown to be the case, it would be an enormous theoretical achievement in its own right. Second, error correction techniques suitable to the boson-sampling model might be developed. This is perhaps the more promising route. Already, steps have been made in overcoming significant experimental scaling problems. Recently, so-called `scattershot' boson-sampling \cite{bib:Scattershot} was presented. This approach overcomes the problem of non-determinism in spontaneous parametric down-conversion sources, allowing a scalable, but still computationally hard device, to be constructed in spite of non-deterministic sources. This discovery is a significant simplification for experimentalists, and overcomes a major scalability issue, a clear demonstration that fundamental scalability issues can be addressed with improved understanding. We hope that future developments in our understanding of boson-sampling will overcome the scalability issues raised here.

Of course, \emph{no} finite sized experiment can ever provide \emph{proof} of an asymptotic statement such as the ECT thesis. However, large-scale experiments \emph{can} provide evidence, if they are combined with an argument that \emph{in principle} the device is arbitrarily scalable. In the case of universal fault-tolerant quantum computation this is possible -- one could in principle demonstrate a large-scale device, and present the theoretical knowledge of fault-tolerance theory that in principle arbitrary scalability is possible, even though the demonstration is of finite size. On the other hand, in the case of boson-sampling, we have presented an argument that arbitrary scalability is problematic with present understanding of scalability. This makes asymptotic claims questionable. However, we hope that future work will further address scalability issues, and, in a similar manner to `scattershot boson-sampling', new developments will overcome these obstacles. Were these fundamental scalability issues addressed, one could make a strong claim that experimental boson-sampling provides evidence against the ECT thesis.

For `conventional' quantum computation, fault-tolerance requires active feedforward, which we is explicitly forbidden in the boson-sampling model. Thus, we expect that fault-tolerance in the boson-sampling model would require entirely different techniques. However, such techniques should remain consistent with the passive-only nature of boson-sampling. Were fault-tolerance techniques developed that deviated from the model, requiring active elements, then one might as well build a universal quantum computer.

%Alternately, it might be shown that the hardness of boson-sampling is inherently robust against certain error models of some magnitude. Were this shown to be the case, the objections raised here might be invalidated.

We emphasise that our claims apply only in the \emph{asymptotic limit} -- we are not making claims about boson-sampling devices of fixed, finite size. With sufficiently low error rates, no doubt larger and larger boson-sampling devices will continue to be demonstrated into the future, and we might even reach the point whereby an experimental implementation demonstrates post-classical capabilities. However, whilst \emph{finite sized} instances of boson-sampling might continue to be demonstrated in the future, we argue that \emph{scalable} boson-sampling in the asymptotic limit is far more challenging, and will require further developments in our understanding of scalability issues. Thus, promoting it as a means by which to disprove the ECT thesis -- inherently an asymptotic question -- is, based on present understanding, questionable.

\begin{acknowledgments}
We thank Scott Aaronson, Jonathan Dowling \& Alexei Gilchrist for helpful discussions. This research was conducted by the Australian Research Council Centre of Excellence for Engineered Quantum Systems (Project number CE110001013), and partly supported by DSTL (contract number DSTLX1000063869).
\end{acknowledgments}

\bibliography{paper}

\begin{thebibliography}{12}
\expandafter\ifx\csname natexlab\endcsname\relax\def\natexlab#1{#1}\fi
\expandafter\ifx\csname bibnamefont\endcsname\relax
  \def\bibnamefont#1{#1}\fi
\expandafter\ifx\csname bibfnamefont\endcsname\relax
  \def\bibfnamefont#1{#1}\fi
\expandafter\ifx\csname citenamefont\endcsname\relax
  \def\citenamefont#1{#1}\fi
\expandafter\ifx\csname url\endcsname\relax
  \def\url#1{\texttt{#1}}\fi
\expandafter\ifx\csname urlprefix\endcsname\relax\def\urlprefix{URL }\fi
\providecommand{\bibinfo}[2]{#2}
\providecommand{\eprint}[2][]{\url{#2}}

\bibitem[{\citenamefont{Aaronson and Arkhipov}(2011)}]{bib:AaronsonArkhipov10}
\bibinfo{author}{\bibfnamefont{S.}~\bibnamefont{Aaronson}} \bibnamefont{and}
  \bibinfo{author}{\bibfnamefont{A.}~\bibnamefont{Arkhipov}},
  \bibinfo{journal}{Proc. ACM STOC (New York)} p. \bibinfo{pages}{333}
  (\bibinfo{year}{2011}).

\bibitem[{\citenamefont{Knill et~al.}(2001)\citenamefont{Knill, Laflamme, and
  Milburn}}]{bib:KLM01}
\bibinfo{author}{\bibfnamefont{E.}~\bibnamefont{Knill}},
  \bibinfo{author}{\bibfnamefont{R.}~\bibnamefont{Laflamme}}, \bibnamefont{and}
  \bibinfo{author}{\bibfnamefont{G.}~\bibnamefont{Milburn}},
  \bibinfo{journal}{Nature} \textbf{\bibinfo{volume}{409}}, \bibinfo{pages}{46}
  (\bibinfo{year}{2001}).

\bibitem[{\citenamefont{Broome et~al.}(2013)\citenamefont{Broome, Fedrizzi,
  Rahimi-Keshari, Dove, Aaronson, Ralph, and White}}]{bib:Broome20122012}
\bibinfo{author}{\bibfnamefont{M.~A.} \bibnamefont{Broome}},
  \bibinfo{author}{\bibfnamefont{A.}~\bibnamefont{Fedrizzi}},
  \bibinfo{author}{\bibfnamefont{S.}~\bibnamefont{Rahimi-Keshari}},
  \bibinfo{author}{\bibfnamefont{J.}~\bibnamefont{Dove}},
  \bibinfo{author}{\bibfnamefont{S.}~\bibnamefont{Aaronson}},
  \bibinfo{author}{\bibfnamefont{T.~C.} \bibnamefont{Ralph}}, \bibnamefont{and}
  \bibinfo{author}{\bibfnamefont{A.~G.} \bibnamefont{White}},
  \bibinfo{journal}{Science} \textbf{\bibinfo{volume}{339}},
  \bibinfo{pages}{6121} (\bibinfo{year}{2013}).

\bibitem[{\citenamefont{Crespi et~al.}(2013)\citenamefont{Crespi, Osellame,
  Ramponi, Brod, Galvao, Spagnolo, Vitelli, Maiorino, Mataloni, and
  Sciarrino}}]{bib:Crespi3}
\bibinfo{author}{\bibfnamefont{A.}~\bibnamefont{Crespi}},
  \bibinfo{author}{\bibfnamefont{R.}~\bibnamefont{Osellame}},
  \bibinfo{author}{\bibfnamefont{R.}~\bibnamefont{Ramponi}},
  \bibinfo{author}{\bibfnamefont{D.~J.} \bibnamefont{Brod}},
  \bibinfo{author}{\bibfnamefont{E.~F.} \bibnamefont{Galvao}},
  \bibinfo{author}{\bibfnamefont{N.}~\bibnamefont{Spagnolo}},
  \bibinfo{author}{\bibfnamefont{C.}~\bibnamefont{Vitelli}},
  \bibinfo{author}{\bibfnamefont{E.}~\bibnamefont{Maiorino}},
  \bibinfo{author}{\bibfnamefont{P.}~\bibnamefont{Mataloni}}, \bibnamefont{and}
  \bibinfo{author}{\bibfnamefont{F.}~\bibnamefont{Sciarrino}},
  \bibinfo{journal}{Nature Phot.} \textbf{\bibinfo{volume}{7}},
  \bibinfo{pages}{545} (\bibinfo{year}{2013}).

\bibitem[{\citenamefont{Tillmann et~al.}(2013)\citenamefont{Tillmann, Daki,
  Heilmann, Nolte, Szameit, and Walther}}]{bib:Tillmann4}
\bibinfo{author}{\bibfnamefont{M.}~\bibnamefont{Tillmann}},
  \bibinfo{author}{\bibfnamefont{B.}~\bibnamefont{Daki}},
  \bibinfo{author}{\bibfnamefont{R.}~\bibnamefont{Heilmann}},
  \bibinfo{author}{\bibfnamefont{S.}~\bibnamefont{Nolte}},
  \bibinfo{author}{\bibfnamefont{A.}~\bibnamefont{Szameit}}, \bibnamefont{and}
  \bibinfo{author}{\bibfnamefont{P.}~\bibnamefont{Walther}},
  \bibinfo{journal}{Nature Phot.} \textbf{\bibinfo{volume}{7}},
  \bibinfo{pages}{540} (\bibinfo{year}{2013}).

\bibitem[{\citenamefont{Spring et~al.}(2013)\citenamefont{Spring, Metcalf,
  Humphreys, Kolthammer, Jin, Barbieri, Datta, Thomas-Peter, Langford, Kundys
  et~al.}}]{bib:Spring2}
\bibinfo{author}{\bibfnamefont{J.~B.} \bibnamefont{Spring}},
  \bibinfo{author}{\bibfnamefont{B.~J.} \bibnamefont{Metcalf}},
  \bibinfo{author}{\bibfnamefont{P.~C.} \bibnamefont{Humphreys}},
  \bibinfo{author}{\bibfnamefont{W.~S.} \bibnamefont{Kolthammer}},
  \bibinfo{author}{\bibfnamefont{X.-M.} \bibnamefont{Jin}},
  \bibinfo{author}{\bibfnamefont{M.}~\bibnamefont{Barbieri}},
  \bibinfo{author}{\bibfnamefont{A.}~\bibnamefont{Datta}},
  \bibinfo{author}{\bibfnamefont{N.}~\bibnamefont{Thomas-Peter}},
  \bibinfo{author}{\bibfnamefont{N.~K.} \bibnamefont{Langford}},
  \bibinfo{author}{\bibfnamefont{D.}~\bibnamefont{Kundys}},
  \bibnamefont{et~al.}, \bibinfo{journal}{Science}
  \textbf{\bibinfo{volume}{339}}, \bibinfo{pages}{798} (\bibinfo{year}{2013}).

\bibitem[{\citenamefont{Aaronson and Arkhipov}(2013)}]{bib:AA13response}
\bibinfo{author}{\bibfnamefont{S.}~\bibnamefont{Aaronson}} \bibnamefont{and}
  \bibinfo{author}{\bibfnamefont{A.}~\bibnamefont{Arkhipov}}
  (\bibinfo{year}{2013}), \eprint{arXiv:1309.7460v2}.

\bibitem[{\citenamefont{Shchesnovich}(2013)}]{bib:Shchesnovich13}
\bibinfo{author}{\bibfnamefont{V.~S.} \bibnamefont{Shchesnovich}}
  (\bibinfo{year}{2013}), \eprint{arXiv:1311.6796}.

\bibitem[{\citenamefont{Shen et~al.}(2013)\citenamefont{Shen, Zhang, and
  Duan}}]{bib:ShenDuan13}
\bibinfo{author}{\bibfnamefont{C.}~\bibnamefont{Shen}},
  \bibinfo{author}{\bibfnamefont{Z.}~\bibnamefont{Zhang}}, \bibnamefont{and}
  \bibinfo{author}{\bibfnamefont{L.-M.} \bibnamefont{Duan}}
  (\bibinfo{year}{2013}), \eprint{arXiv:1310.4860}.

\bibitem[{\citenamefont{Tichy et~al.}(2013)\citenamefont{Tichy, Mayer,
  Buchleitner, and M{\o}lmer}}]{bib:Molmer13}
\bibinfo{author}{\bibfnamefont{M.~C.} \bibnamefont{Tichy}},
  \bibinfo{author}{\bibfnamefont{K.}~\bibnamefont{Mayer}},
  \bibinfo{author}{\bibfnamefont{A.}~\bibnamefont{Buchleitner}},
  \bibnamefont{and} \bibinfo{author}{\bibfnamefont{K.}~\bibnamefont{M{\o}lmer}}
  (\bibinfo{year}{2013}), \eprint{arXiv:1312.3080}.

\bibitem[{\citenamefont{Shchesnovich}(2014)}]{bib:shchesnovich2014conditions}
\bibinfo{author}{\bibfnamefont{V.}~\bibnamefont{Shchesnovich}},
  \bibinfo{journal}{arXiv:1403.4459}  (\bibinfo{year}{2014}).

\bibitem[{\citenamefont{Lund et~al.}(2013)\citenamefont{Lund, Laing,
  Rahimi-Keshari, Rudolph, O'Brien, and Ralph}}]{bib:Scattershot}
\bibinfo{author}{\bibfnamefont{A.~P.} \bibnamefont{Lund}},
  \bibinfo{author}{\bibfnamefont{A.}~\bibnamefont{Laing}},
  \bibinfo{author}{\bibfnamefont{S.}~\bibnamefont{Rahimi-Keshari}},
  \bibinfo{author}{\bibfnamefont{T.}~\bibnamefont{Rudolph}},
  \bibinfo{author}{\bibfnamefont{J.~L.} \bibnamefont{O'Brien}},
  \bibnamefont{and} \bibinfo{author}{\bibfnamefont{T.~C.} \bibnamefont{Ralph}}
  (\bibinfo{year}{2013}), \eprint{arXiv:1305.4346}.

\end{thebibliography}

\end{document}